\documentstyle [twocolumn,prl,aps] {revtex}
\begin{document}
\draft
\begin{title}
{Electronic entropy, shell structure, and size-evolutionary 
patterns of metal clusters}
\end{title} 
\author{Constantine Yannouleas and Uzi Landman }
\address{
School of Physics, Georgia Institute of Technology,
Atlanta, Georgia 30332-0430 }
\date{Phys. Rev. Lett. 78, 1424 [24 February 1997]}
\maketitle
\begin{abstract}
We show that electronic-entropy effects in the size-evolutionary patterns of 
relatively small (as small as 20 atoms), simple-metal clusters become 
prominent already at moderate temperatures. Detailed agreement between our 
finite-temperature-shell-correction-method calculations and experimental 
results is obtained for certain temperatures. This agreement includes a 
size-dependent smearing out of fine-structure features, accompanied by a 
measurable reduction of the heights of the steps marking major-shell and 
subshell closings, thus allowing for a quantitative analysis of cluster 
temperatures. 
\end{abstract}
~~~~\\

\narrowtext

Since the discovery \cite{knig} of electronic-shell-structure 
features in the abundance spectra of sodium clusters, similar features 
(the major ones
corresponding to the degeneracies of a spherically symmetric mean-field
potential \cite{ekar,beck}) have been routinely observed \cite{dehe} in the 
size-evolutionary 
patterns (SEP's) associated with other single-particle properties of both 
alkali- and noble-metal clusters. Specifically, such patterns pertain to 
ionization potentials (IP's) \cite{saun,jack,alam}, electron affinities (EA's)
\cite{eato,tayl}, monomer separation energies (MSE's) \cite{brec1}, and 
fission dissociation energies \cite{brec3}.

It was early realized \cite{clem} that the secondary features in the mass 
spectra required consideration of deformed cluster shapes \cite{saun,clem}.
When triaxial (ellipsoidal) shapes were considered in the framework of
Shell Correction Methods (SCM) \cite{yann0,yann1,yann2,yann4},
substantial overall systematic agreement was achieved 
\cite{yann1,yann2} between theory and experimental observations pertaining to 
the major and the fine structure of the aforementioned SEP's.

While deformation effects have been extensively studied, an understanding of 
the physical origins of thermal effects and their relation to the SEP's is 
still lacking, even though the experiments are necessarily made with clusters 
at finite temperatures. Moreover, experimental determination of cluster
temperatures remains a challenge, motivating the development of theoretical
methods capable of extracting such information. 

While thermodynamic entropic contributions associated with the ionic degrees
of freedom can be obtained from first-principles molecular-dynamics
simulations \cite{andr,land}, or from considerations of 
shape fluctuations in simpler models
\cite{akul,mann,note3}, for sufficiently large simple-metal clusters
$M_N$ with $N > 20$, the electronic entropy (which has not as yet been
included in such studies) dominates the thermal characteristics of the
shell structure, even at moderate temperatures. The prominent thermal
effects associated with the electronic entropy are the focus of this paper.  

Without the simultaneous consideration of shape fluctuations, 
thermal effects pertaining to the electronic degrees of freedom 
and their relevance for the understanding
of mass abundance spectra have been considered in the case of 
spherical neutral sodium clusters \cite{brac},
and in a recent report \cite{frau2} on the thermodynamics of neutral 
sodium clusters with axially symmetric shapes. In contrast to our findings, 
these studies have suggested that electronic-entropy effects at moderate 
temperatures are not important for clusters with less than 100 atoms. 

The theoretical method used in this paper is a 
finite-temperature (FT) --SCM developed by us, which incorporates all three  
of the aforementioned aspects, namely, triaxial 
deformations, entropy of the electrons, 
and thermal effects originating from shape fluctuations. 
Furthermore, through a direct comparison with experimental measurements, we 
demonstrate that this method  can be employed for determining cluster 
temperatures.

Since the number of delocalized valence electrons is fixed for a given
cluster, $M_N^{x \pm}$, we need to use \cite{dent,kubo} the {\it canonical\/} 
ensemble in calculating their thermodynamical properties. For determining
the free energy, $F(\beta,N,x)$ ($\beta=1/T)$, which incorporates the 
electronic entropy, we separate it, in analogy with the
zero-temperature limit \cite{yann1}, into a smooth, liquid-drop-model part, 
$\widetilde{F}_{\text{LDM}}$, and a Strutinsky-type \cite{stru} 
shell-correction term, $\Delta F_{\text{sh}}$. 
The shell correction term is specified as the difference
$\Delta F_{\text{sh}}=F_{\text{sp}}-\widetilde{F}_{\text{sp}}$, where $F_{\text{sp}}$ is the canonical free energy of the valence electrons viewed 
as independent single particles in their effective mean-field potential. 
For calculating the canonical $F_{\text{sp}}$, we
adopt a number-projection method \cite{ross}, according to which 
\[ 
F_{\text{sp}} = -\frac{1}{\beta} \ln \{ \int_0^{2 \pi} 
\frac{d\phi}{2 \pi} Z_{\text{GC}} (\beta, \beta \mu + i \phi)
e^{-(\beta \mu + i \phi) N_e} \}~,
\] 
where $N_e$ is the number of electrons and $\mu$ is the chemical potential
of the equivalent grand-canonical ensemble. The grand-canonical 
partition function, $Z_{\text{GC}}$, is given by 
\[
Z_{\text{GC}}(\beta, \beta \mu) = 
\prod_{i} (1+ e^{-\beta(\varepsilon_i - \mu)})~,
\]
where $\varepsilon_i$ are the single-particle levels of the modified
Nilsson hamiltonian pertaining to triaxial shapes \cite{yann1}. 
The temperature-dependent average 
$\widetilde{F}_{\text{sp}}$ \cite{jens} is specified using the same
expressions as for $F_{\text{sp}}$, but with a set of smooth levels 
\cite{frau2} defined as
$\widetilde{\varepsilon}_i = 
\widetilde{E}^{\text{osc}}_{\text{sp}}(i) -
\widetilde{E}^{\text{osc}}_{\text{sp}}(i-1)$, where 
$\widetilde{E}^{\text{osc}}_{\text{sp}}(N_e)$
is the zero-temperature Strutinsky average
of the single-particle spectrum of an 
anisotropic, triaxial oscillator (see section II.C. of Ref.\ \cite{yann1}).   

The LDM term $\widetilde{F}_{\text{LDM}}$ consists of three contributions;
a volume, a surface, and a curvature term. Since volume conservation during 
deformation is assumed, we need not consider the temperature dependence of 
the corresponding term when calculating observables, such as IP's and MSE's,
associated with processes which do not change the total number of atoms, $N$.
The experimental temperature dependence of the surface tension, 
$\sigma (T) = c_0 - c_1 (T-T_{\text{mp}})$, is taken from standard
tables \cite{hand} ($T_{\text{mp}}$ are melting-point temperatures 
\cite{elem}), but normalized to yield the $\sigma(T=0)$ value used 
in our earlier calculations \cite{yann1}. 
Since no experimental information is available regarding the
curvature coefficient, $A_c$, we assume the same relative temperature
dependence as for $\sigma (T)$ normalized to the $T=0$ value used earlier
\cite{yann1}. Finally, the expansion of the Wigner-Seitz radius due to the
temperature is determined according to the cofficient of linear thermal
expansion \cite{elem}. With these modifications, the remaining steps in the
calculation of $\widetilde{F}_{\text{LDM}}$ follow closely section II.A. of
Ref.\ \cite{yann1} (for simplicity, the work function $W$ is assumed 
temperature independent).
 
According to the general theory of thermal fluctuating phenomena \cite{landau},
the triaxial shapes of the clusters, specified
by the $\beta_H$ and $\gamma_H$ Hill-Wheeler parameters \cite{hw}, will
explore \cite{yann3} the free-energy surface,
$F(\beta, N, x; \beta_H, \gamma_H)$, with a 
probability,
\[
P(\beta_H, \gamma_H) = 
{\cal Z}^{-1} \exp[- \beta F(\beta, N, x; \beta_H, \gamma_H)]~,
\]
the quantity ${\cal Z}$ being the classical Boltzmann-type partition
function,
\[
{\cal Z} = \int d\tau \exp[- \beta F(\beta, N, x; \beta_H, \gamma_H)]~,
\]
and $d \tau = \beta_H^4 |\sin(3\gamma_H)|d\beta_H d\gamma_H$ the proper
volume element \cite{bm}.
Thus, finally, the free energy, $\langle F(\beta, N, x) \rangle$, 
averaged over the shape fluctuations can be written as
\[
\langle F(\beta, N, x) \rangle =
\int d \tau F(\beta, N, x; \beta_H, \gamma_H) P(\beta_H, \gamma_H)~.
\]

We will present results pertaining to IP's and MSE's 
($M_N^+ \rightarrow M_{N-1}^+ + M$), which at finite 
temperatures are defined as,
\[
I_N = 
\langle F(\beta, N, x=+1) \rangle -
\langle F(\beta, N, x=0) \rangle~,
\]
and
\begin{eqnarray}
D_{\text{1,N}}^+ = 
\langle F && (\beta, N-1, x=+1) \rangle +
\langle F(\beta, N=1, x=0) \rangle \nonumber \\
&& - \langle F(\beta, N, x=+1) \rangle~, \nonumber
\end{eqnarray}
respectively.

Our calculations (solid dots) for the IP's of K$_N$ clusters for three 
temperatures, $T=10$ K, 300 K, and 500 K,
are displayed in Fig.\ 1, and are compared with the experimental measurements
\cite{saun} (open squares; the experimental uncertainties are 
0.06 eV for $N \leq 30$ and 0.03 eV for $N > 30$). 
As was the case with our earlier $T=0$ K results
\cite{yann1}, the $T=10$ K theoretical results exhibit the following 
two characteristics: (i) Above $N=21$, a 
pronounced fine structure between major-shell closures which is not present in
the experimental measurements; (ii) Steps at the major-shell closures which
are much larger than the experimental ones \cite{note5} (three-to-five times 
for $N=40$, 58, and 92, and two-to-three times for $N=8$ and 20).

The agreement between theory and experiment is significantly improved at
$T=300$ K. Indeed, in comparison with the lower-temperature
calculations, the $T=300$ K
results exhibit the following remarkable changes: (i) Above $N=21$, the
previously sharp fine-structure features are smeared out, and as a 
result, the theoretical curve follows closely the smooth modulations of the  
experimental profile. In the size range $N=21-34$, three rounded, 
hump-like formations (ending to the right at the subshell closures at $N=26$,
30, and 34) survive in very good agreement with the experiment (the sizes of 
the drops at $N=26$, 30 and 34 are comparable to the experimental ones
\cite{note}); (ii) The sizes of the IP drops at $N=20$, 40, 58, and 92
are reduced drastically and are now comparable to the experimental ones.
In the size range $N \leq 20$, the modifications are not as dramatic. Indeed,
one can clearly see that the pattern of odd-even alternations
remains well defined, but with a moderate attenuation in amplitude, again 
in excellent agreement with the experimental observation. 

For $T=500$ K, the smearing out of the shell structure  
progresses even further, obliterating the agreement between theory and 
experiment. Specifically, the steps at the subshell closures at $N=26$ and
30, as well as at the major-shell closures at $N=40$, 58, and 92 are rounded
and smeared out over several clusters
(an analogous behavior has been observed in the logarithmic abundance spectra
of hot, singly cationic, copper, silver, and gold clusters \cite{kata}).
At the same time, however, the odd-even alternation remains well defined
for $N \leq 8$. We further notice that, while some residue of fine structure 
survives in the range $N=9-15$, the odd-even alternations there are 
essentially absent (certain experimental measurements \cite{kapp} of the IP's 
of hot Na$_N$ clusters appear to conform to this trend).

To ascertain the relative weight of the two thermal processes incorporated
in our FT-SCM, we display in the middle panel of Fig.\ 1 (uppermost curve with
open circles) the theoretical IP's at $T=300$ K in the case when the electronic
entropy is neglected. A comparison with the results (solid dots) when both 
electronic and shape-fluctuations entropic contributions are included  
demonstrates the principal role of the electronic entropy in shaping the 
thermal effects of the SEP's.

Fig.\ 2 displays for two temperatures (10 K and 300 K) 
the FT-SCM results (solid dots) for the MSE's associated with the
K$_N^+$ clusters, along with available experimental measurements
\cite{brec1} (open squares) in the size range $N=4-23$. 
Compared to the $T=10$ K results, the theoretical results at 
$T=300$ K are in better agreement with the experimental ones
due to an attenuation of the amplitude of the alternations
(e.g., notice  the favorable reduction in the size of the drops at
$N=9$, 15, and 21). In spite of this amplitude attenuation, it is remarkable 
that the $T=300$ K SCM results in this size range preserve in 
detail the same relative pattern as the $T=10$ K ones (in particular, 
the well-defined odd-even oscillations in the range $N=4-15$ and the
ascending quartet at $N=16-19$ followed by a dip at $N=20$).

As a last example, Fig.\ 3 displays for three temperatures (10 K, 
800 K, and 2000 K) our FT-SCM results for the IP's of 
Ag$_N$ clusters, and compares them with available experimental 
measurements \cite{jack,note2}.
Again, going from the $T=10$ K to the $T=800$  results, 
we observe that an attenuation of the amplitude of alternations brings
theory and experiment in better agreement (e.g., in the latter case, the sizes
of the theoretical IP drops at $N=6$, 8, 14, 20, 26, 30, and 34 are comparable
to the experimental ones). Finally, at $T=2000$ K, both major and
fine structure tend to vanish for $N > 8$.

In conclusion, we showed that the SEP's of the single-particle properties of 
simple-metal clusters are governed by the interplay of ellipsoidal 
deformations and thermodynamics (entropy) of the electronic degrees of 
freedom, while the entropic contribution of shape fluctuations plays a smaller
role \cite{note6}. We further demonstrated that
electronic-entropy effects are reflected in prominent experimental 
signatures already at moderate temperatures and for relatively small
sizes. This behavior, which had not been previously understood in the case 
of metal clusters, correlates with an order-of-magnitude estimate \cite{bm} 
from theoretical nuclear physics. Accordingly,
the shell-correction term in the free energy depends on $T$ as 
$\Delta F_{\text{sh}}(T) = \Delta F_{\text{sh}}(0) \tau/\sinh \tau$, which for 
$\tau > 1$ takes the form $2\Delta F_{\text{sh}}(0) \tau \exp(-\tau)$ with 
$\tau=2\pi^2 T/ (\hbar \omega_{\text{sh}})$, with $\hbar \omega_{\text{sh}}$ 
being the energy spacing between shells. Because of the large $2\pi^2$ factor, 
the shell-structure effects decrease rapidly, even for 
temperatures that are only a fraction of the shell spacing 
(see p. 608 of Ref.\ \cite{bm}). Thus, the shell structure 
``melts'' for temperatures as low as $T/(\hbar \omega_{\text{sh}})=0.25$
\cite{note4}. The FT-SCM calculations presented here 
correspond to temperature ranges well below this ratio, where analysis of
the extent of attenuation of shell-structure signatures allows determination 
of cluster temperatures.

Compared to nuclear and atomic physics, which have heretofore provided the 
prototypes for fermionic shell structure, the SEP's of metal clusters stand 
apart, since the experimentally available SEP's of nuclei 
(e.g., those of neutron separation 
energies \cite{bm2}) and of atoms (i.e., IP's \cite{bm2}) correspond 
strictly to zero-temperature. Finally, it is natural to conjecture 
that thermal (in particular, electronic-entropy) effects will influence the 
height of fission barriers and fragmentation channels \cite{brec4}.

This research is supported by the US Department of Energy (Grant No.
FG05-86ER-45234).

\begin{figure}
\caption{
IP's of K$_N$ clusters at three temperatures, $T=10$ K, 300 K, and 500 K.
Solid dots: Theoretical FT-SCM results. Open squares: experimental 
measurements \protect\cite{saun}. The uppermost curve (open circles) in the
middle panel displays theoretical results when the electronic entropy is 
neglected. The $y$-scale in this instance is to the right. 
}
\end{figure}
\begin{figure}
\caption{
MSE's of K$_N^+$ clusters at two temperatures, $T=10$ K, and 300 K.
Solid dots: Theoretical FT-SCM results. Open squares: experimental 
measurements \protect\cite{brec1}. 
To facilitate comparison, the SCM results 
at the higher temperature have been shifted by 0.07 eV, so that the 
theoretical curves at both temperatures refer to the same point at $N=10$.
}
\end{figure}
\begin{figure}
\caption{
IP's of Ag$_N$ clusters at three temperatures, $T=10$ K, 800 K, and 2000 K.
Solid dots: Theoretical FT-SCM results. Open squares: experimental 
measurements \protect\cite{jack}.
}
\end{figure}


\begin{references}
\bibitem{note5}
In contrast, experimental IP's for cold Na$_N$ clusters
are in good overall agreement with our $T=0$ K SCM results regarding both
characteristics (see Refs.\ \cite{yann1,yann2}).
\bibitem{note}
Notice that experimental measurements at $N=33$ and 35 have not been obtained
(see Ref.\ \cite{saun}).
\bibitem{note2}
At large $N$, the quoted experimental values appear to converge to the
value of an effective work function, $W_{\text{eff}}$, which is by $\sim 0.8$ 
eV systematically higher than the measured polycrystalline one ($W=4.26$ eV),
which was used in our calculations. Indeed, the low-temperature 
measurements for the IP's of Ag$_N$ clusters given in Ref.\ \cite{alam}
converge to $W=4.26$ eV.
\bibitem{note6}
In other instances, e.g., the widths of collective plasma excitations,
shape fluctuations may play a larger role (see Ref.\ \cite{yann3}).
\end{references}
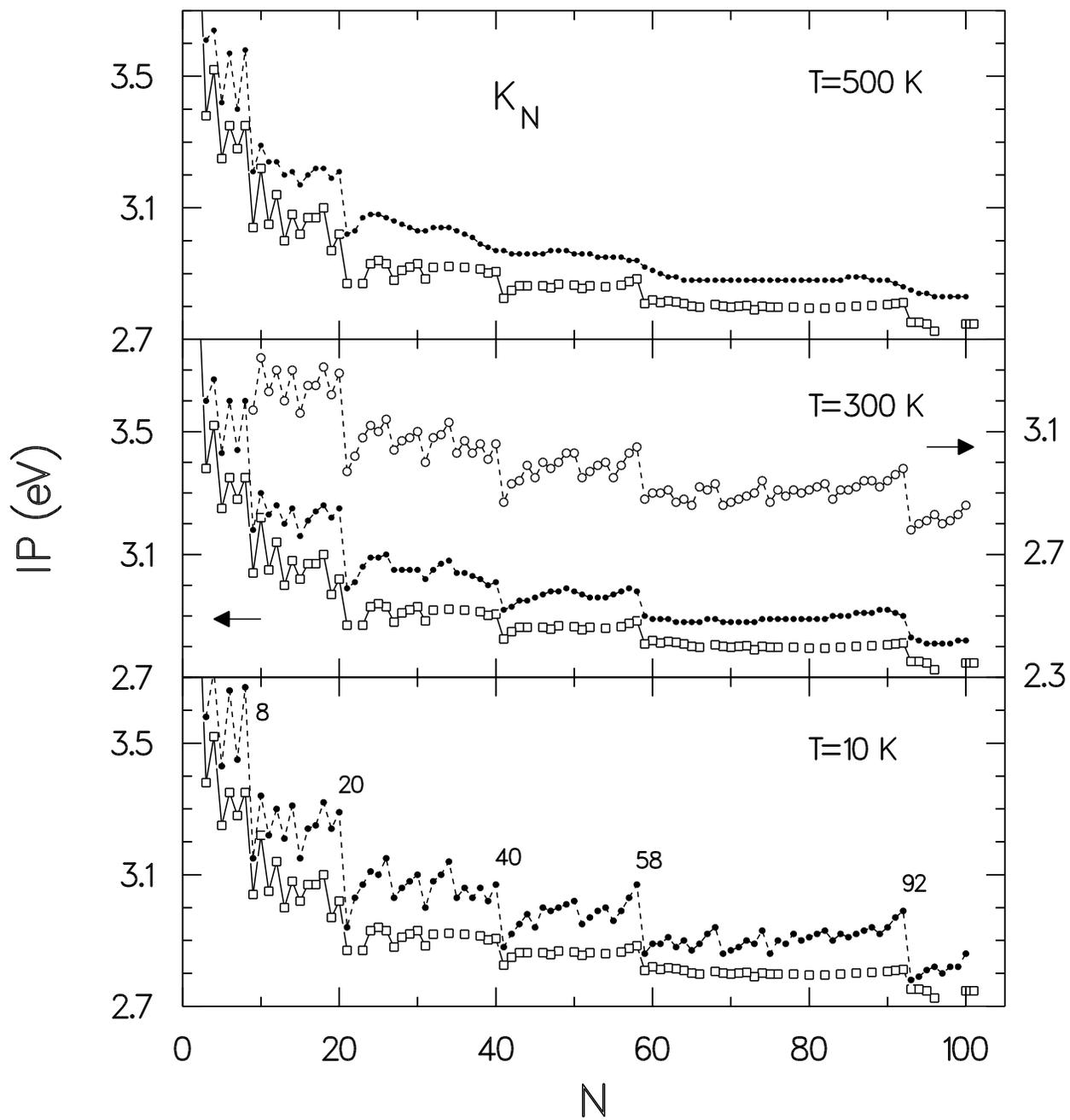
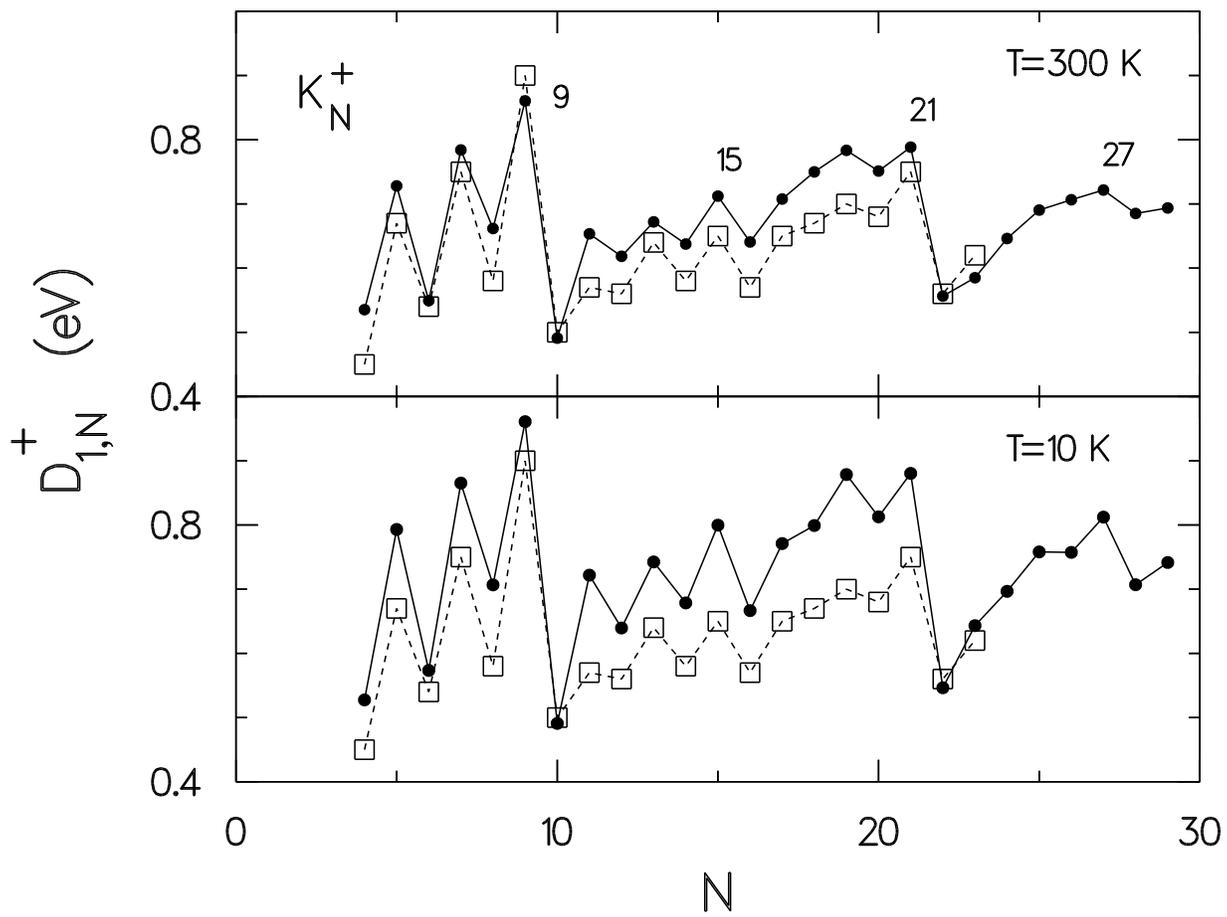
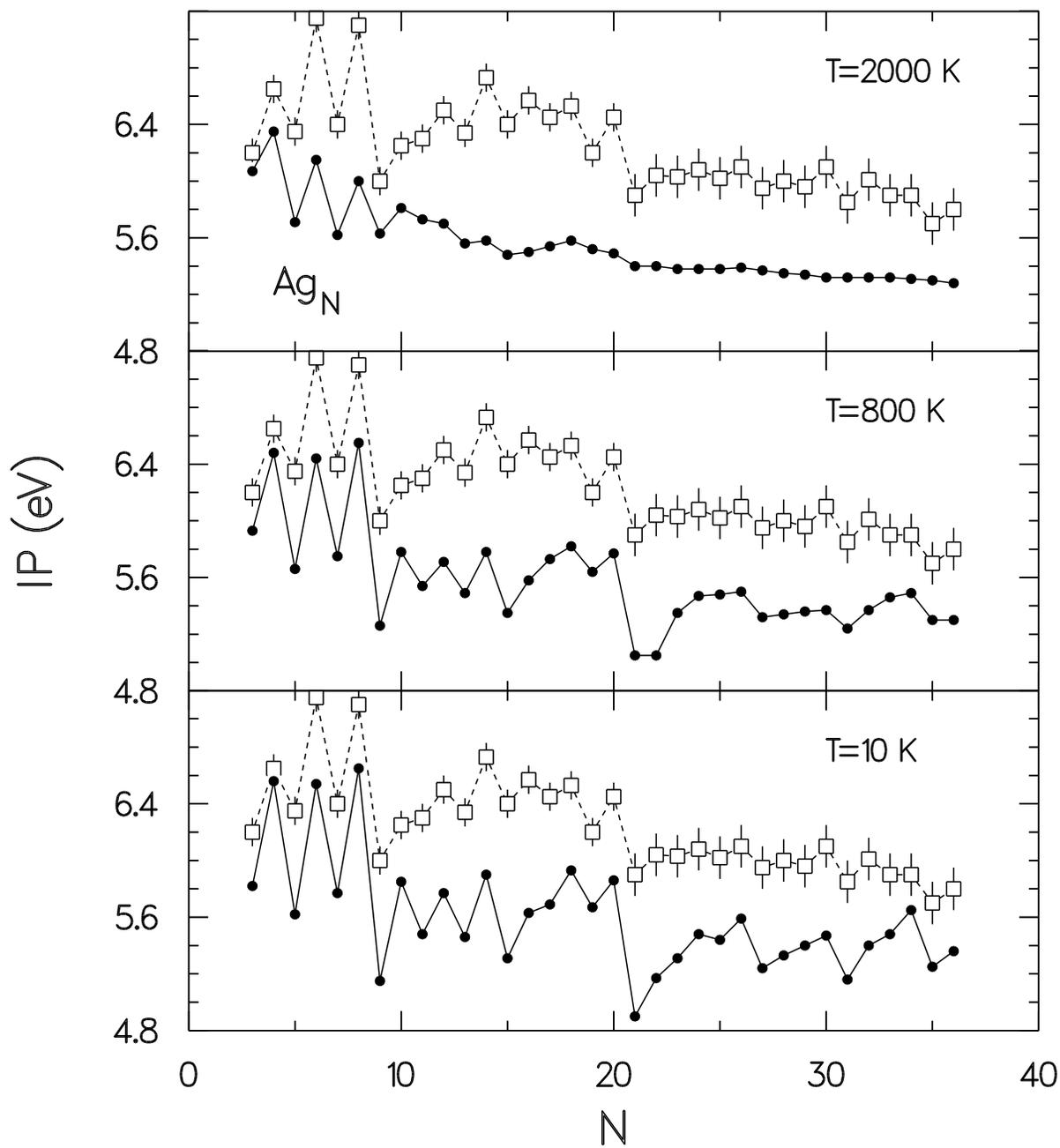
\end{document}